\documentclass[draftclsnofoot,11pt,onecolumn]{IEEEtran}
\usepackage{latexsym}
\usepackage{amsmath}
\usepackage{amsfonts}
\usepackage{amssymb}
\usepackage{amsthm}
\usepackage{cite}
\usepackage{graphicx} % use this when importing ps- and eps-files
\usepackage{subfigure} % without it, the subfigure may not exihibit
\usepackage{color}
\usepackage{algorithm} %format of the algorithm
\usepackage{algorithmic} %format of the algorithm
\usepackage{bm}
\usepackage{longtable}% for long tables
\usepackage{hyperref}
\usepackage{graphicx}  % Used to add the photoes

\newtheorem{theorem}{Theorem}
\newtheorem{definition}{Definition} % Let definition show the same serial number with theorem
\newtheorem{lemma}{Lemma} % Let lemma show the same serial number with theorem
\newtheorem{remark}{Remark} % Let remark show the same serial number with theorem
\ifCLASSINFOpdf
\else
\fi
\begin{document}
\title{Deterministic Analysis of Weighted BPDN With Partially Known Support Information}

\author{

\thanks{
This work was partially supported by the China Postdoctoral Science Foundation under Grant 2018M643390 and the Natural Science Foundation of China under Grant 61673015. (\emph{Corresponding author: Jianjun Wang.})
}

Wendong Wang and Jianjun Wang

\thanks{W. Wang and J. Wang are with the School of Mathematics and Statistics, Southwest University, Chongqing 400715, China (e-mail: wdwang@swu.edu.cn; wjj@swu.edu.cn).}

}

\maketitle

\begin{abstract}
In this paper, with the aid of the powerful Restricted Isometry Constant (RIC), a deterministic (or say non-stochastic) analysis, which includes a series of sufficient conditions (related to the RIC order) and their resultant error estimates, is established for the weighted Basis Pursuit De-Noising (BPDN) to guarantee the robust signal recovery when Partially Known Support Information (PKSI) of the signal is available. Specifically, the obtained conditions extend nontrivially the ones induced recently for the traditional constrained weighted $\ell_{1}$-minimization model to those for its unconstrained counterpart, i.e., the weighted BPDN. The obtained error estimates are also comparable to the analogous ones induced previously for the robust recovery of the signals with PKSI from some constrained models. Moreover, these results to some degree may well complement the recent investigation of the weighted BPDN which is based on the stochastic analysis.
\end{abstract}


% \fi
\IEEEpeerreviewmaketitle

\section{Introduction}\label{Section1}
\IEEEPARstart{C}{ompressed/compressive} Sensing (CS), see, e.g., \cite{RIP-notation,DL-Donoho-C-IEEE-IT-2006,CS-textbook}, has captured a lot of attention of the researchers in a wide range of fields over the past decade. In CS, one gets the observations of signal $\widehat{\bm{x}}\in\mathbb{R}^{n}$ via the following model
\begin{align}\label{b=ax+z}
\bm{b}=A\widehat{\bm{x}}+\bm{z},
\end{align}
where $A\in\mathbb{R}^{m\times n}(m\ll n)$ is called the measurement matrix and $\bm{z}\in\mathbb{R}^{m}$ denotes the additive noise that satisfies a certain constraint. One of the key goals of CS is to effectively recover the original signal $\widehat{\bm{x}}$ based on $A$ and $\bm{b}$. It has been shown that, if $\widehat{\bm{x}}$ is $k$-sparse with $k<m$ and $A$ satisfies certain conditions related to $k$, see, e.g., \cite{RIP-Recovery-conditions-04142,RIP-Recovery-conditions-tk-1,RIP-Recovery-conditions-tk-2,RIP-Recovery-conditions-2k-Lp,Coherence-Recovery-conditions-sharp}, then one can achieve this goal by solving an $\ell_{1}$-minimizing problem, i.e.,
\begin{align}\label{Constrained-L1}
\min_{\bm{x}\in\mathbb{R}^{n}}\|\bm{x}\|_{1}~~s.t.~~\|\bm{b}-A\bm{x}\|_{2}\leq\epsilon,
\end{align}
where $\epsilon\geq0$ represents for the noise level, and we take $\epsilon=0$ if there is no noise, i.e., $\bm{z}=0$.

The above $\ell_{1}$-minimization approach has been demonstrated to be effective in robust signal recovery. However, it does not incorporate any prior information on signal support since the $\ell_{1}$-norm treats the entries of variable $\bm{x}$ equally. In fact in many practical applications such as the time-series signal processing, see, e.g., \cite{CS-prior-ISIT-2009-Khajehnejad,CS-prior-ISIT-2009-Vaswani,L-Jacques-SP-2010}, it is often possible to estimate a part of the signal support information. It thus becomes very necessary and important to use such prior information to further enhance the recovery performance of \eqref{Constrained-L1}. This consideration directly leads to the following weighted $\ell_{1}$-minimization problem
\begin{align}\label{Constrained-WL1}
\min_{\bm{x}\in\mathbb{R}^{n}}\|\bm{x}\|_{\bm{w},1}\triangleq\sum_{i=1}^{n}\bm{w}_{i}|\bm{x}_{i}|~~s.t.~~\|\bm{b}-A\bm{x}\|_{2}\leq\epsilon,
\end{align}
where $\bm{w}\triangleq[\bm{w}_{1},\cdots,\bm{w}_{n}]^{T}$ denote the weights. For simplicity, in this paper we only consider a binary choice of $\bm{w}$, i.e.,
\begin{align*}
\begin{split}
\bm{w}_{i}=\left\{
  \begin{array}{ll}
    w\in[0,1], & i\in K\\
    1, & i\in K^{c}
  \end{array}
\right.,
\end{split}
\end{align*}
where $K\subset[n]\triangleq\{1,2,\cdots, n\}$ is a given set, which models the Partially Known Support Information (PKSI) of $\widehat{\bm{x}}$. This problem has been well investigated in the past few years, see, e.g., \cite{CS-prior-TIT-2012,ince2013nonconvex,zhan2015time,flinth2016optimal,CS-prior-NSP-Mansour-2017,needell2017weighted,CS-prior-SP-CHen-2018,
CS-prior-ACHA-CL-In-Press}. It was proved by Friedlander, \emph{et al.} in \cite{CS-prior-TIT-2012} that if $K$ includes half of the accurate support of $\bm{\widehat{x}}$ at least, then \eqref{Constrained-WL1} will perform robustly under much weaker conditions than the analogous ones for \eqref{Constrained-L1}. In \cite{flinth2016optimal}, Flinth studied the optimal choice for general weights. Recently, Chen, \emph{et al.} in \cite{CS-prior-SP-CHen-2018} and \cite{CS-prior-ACHA-CL-In-Press} obtained some much tighter conditions for \eqref{Constrained-WL1}, and these conditions were proved to be sharp when the desired signal $\bm{\widehat{x}}$ is exactly sparse and is also measured without noise.

%Recent studies, see, e.g., \cite{IRLS-Lq-2013}, shown that an unconstrained model may behave much better than its constrained counterpart in dealing with the noisy measurements and the approximately sparse signals. Motivated by this fact,

In this paper, we consider the robust recovery of the signals with PKSI via the weighted Basis Pursuit De-Noising (BPDN)
\begin{align}\label{UnConstrained-WL1}
\min_{\bm{x}\in\mathbb{R}^{n}}\|\bm{x}\|_{\bm{w},1}+\frac{1}{2\lambda}\|\bm{b}-A\bm{x}\|_{2}^{2},
\end{align}
where $\lambda$ is a positive parameter. Obviously, \eqref{UnConstrained-WL1} will be reduced to the widely known BPDN if one sets $w=1$ (i.e., no support information is available). Although there exists a large amount of research on the BPDN, see, e.g., \cite{BPDN-2004,BPDN-2005,BPDN-2008-Zhu,BPDN-2014-Lin,BPDN-2015-Shen,BPDN-2018-Submitted,BPDN-2019-JCAM,CoherenceWang-2018-submitted}, the theoretical analysis of \eqref{UnConstrained-WL1} for sparse recovery is relatively less studied. We note that Lian, \emph{et al.} recently studied \eqref{UnConstrained-WL1} from both theoretical and experimental aspects in \cite{lian2018weighted}, where they called it weighted LASSO. However, their obtained results are established on the stochastic strategy, and they are totally different from ours that are established in a deterministic manner.

The main contribution of this paper is that a series of (tight) sufficient conditions as well as their resultant error estimates are established for \eqref{UnConstrained-WL1} with the help of the Restricted Isometry Property (RIP) \cite{RIP-notation}, which to some degree well complement the recent theoretical analysis of the weighted BPDN (see, \cite{lian2018weighted}) that is based on the stochastic strategy.

%The rest of this paper is organized as follows. In Section \ref{Notations-Preliminaries} we introduce some notations and key lemmas. We present our main theoretical results in Section \ref{Main-results}, and then conclude the paper in Section \ref{conclusion-part}.

\section{Notations and Preliminaries}
\label{Notations-Preliminaries}
In this section, we first introduce some basic notations. For any given index set $S\subset[n]$, we denote $\bm{h}_{S}$ as a vector whose entries $(\bm{h}_{S})_{i}=\bm{h}_{i}$ for $i\in S$ and 0 otherwise, and also denote the best $s$-term approximate $\bm{x}_{\max(s)}$ of any signal $\bm{x}\in\mathbb{R}^{n}$ as
\begin{align*}
\bm{x}_{\max(s)}=\mathop{\arg\min}_{\|\bm{y}\|_{0}\leq s}\|\bm{y}-\bm{x}\|_{2}.
\end{align*}
\begin{definition}
A matrix $A\in\mathbb{R}^{m\times n}$ is said to obey the RIP of order $k$, if there exists a constant $\delta\in(0,1)$ such
\begin{align}\label{RIP}
(1-\delta)\|\bm{x}\|_{2}^{2}\leq\|A\bm{x}\|_{2}^{2}\leq(1+\delta)\|\bm{x}\|_{2}^{2}
\end{align}
for every $k$-sparse signal $\bm{x}\in\mathbb{R}^{n}$. The smallest positive $\delta$ that satisfies \eqref{RIP} is denoted by $\delta_{k}$ \footnote{When $k$ is not an integer, we define $\delta_{k}$ as $\delta_{\lceil k\rceil}$.} and is known as the Restricted Isometry Constant (RIC).
\end{definition}

We also need the following two lemmas.
\begin{lemma}\label{NP-Lemma1}
Assume that $E,K\subseteq[n]$ are two sets with $|E|=k$, $|K|=\rho k$ and $|E\cap K|=\alpha \rho k$ for some $\rho\geq0$ and $0\leq\alpha\leq1$, and define
\begin{align}\label{d-defn}
\begin{split}
d=\left\{
  \begin{array}{ll}
    1-\alpha\rho+\max\{\alpha,1-\alpha\}\rho, & 0\leq w<1\\
    1, & w=1
  \end{array}
\right.
\end{split}.
\end{align}
If $\bm{b}$ is observed through \eqref{b=ax+z} with the noise constrain $\|\bm{z}\|_{2}\leq\epsilon$, then for the optimal solution
$\bm{x}^{\sharp}$ of \eqref{UnConstrained-WL1}, we have
\begin{align}\label{NP-Lemma1-results1}
\|A\bm{h}\|_{2}^{2}-2\epsilon\|A\bm{h}\|_{2}\leq4\lambda\left(w\|\bm{\widehat{x}}_{E^{c}}\|_{1}+(1-w)\|\bm{\widehat{x}}_{K^{c}\cap E^{c}}\|_{1}\right)+2\theta\sqrt{k}\lambda\|\bm{h}_{\max(dk)}\|_{2}-2\lambda\|\bm{h}_{E^{c}}\|_{1}
\end{align}
and
\begin{align}\label{NP-Lemma1-results2}
\|\bm{h}_{E^{c}}\|_{1}\leq2(w\|\bm{\widehat{x}}_{E^{c}}\|_{1}+(1-w)\|\bm{\widehat{x}}_{K^{c}\cap E^{c}}\|_{1})+\theta\sqrt{k}\|\bm{h}_{\max(dk)}\|_{2}+\frac{\epsilon}{\lambda}\|A\bm{h}\|_{2},
\end{align}
where $\bm{h}=\bm{x}^{\sharp}-\bm{\widehat{x}}$ and $\theta$ is denoted by \eqref{theta-defn}.
\end{lemma}

\begin{IEEEproof}
Since $\bm{x}^{\sharp}$ is the optimal solution of \eqref{UnConstrained-WL1}, we have
\begin{align*}
\|\bm{x}^{\sharp}\|_{\bm{w},1}+\frac{1}{2\lambda}\|\bm{b}-A\bm{x}^{\sharp}\|_{2}^{2}\leq\|\bm{\widehat{x}}\|_{\bm{w},1}+\frac{1}{2\lambda}\|\bm{b}-A\bm{\widehat{x}}\|_{2}^{2},
\end{align*}
which is equivalent to
\begin{align}\label{Unconstrained-L1-L2Noise-Estimate-LR}
\|A\bm{h}\|_{2}^{2}-2\langle \bm{z}, A\bm{h}\rangle\leq& 2\lambda(\|\bm{\widehat{x}}\|_{\bm{w},1}-\|\bm{x}^{\sharp}\|_{\bm{w},1}).
\end{align}
As to the left-hand side of \eqref{Unconstrained-L1-L2Noise-Estimate-LR}, we have
\begin{align}\label{Unconstrained-L1-L2Noise-Estimate-L}
\text{LHS}\geq\|A\bm{h}\|_{2}^{2}-2\epsilon\|A\bm{h}\|_{2}.
\end{align}
As to the right-hand side of \eqref{Unconstrained-L1-L2Noise-Estimate-LR}, we know from \cite{CS-prior-TIT-2012} that
\begin{align}\label{new-added-1}
\frac{\text{RHS}}{2\lambda}\leq& 2\left(w\|\bm{\widehat{x}}_{E^{c}}\|_{1}+(1-w)\|\bm{\widehat{x}}_{K^{c}\cap E^{c}}\|_{1}\right)+w\|\bm{h}_{E}\|_{1}+(1-w)\|\bm{h}_{U}\|_{1}-\|\bm{h}_{E^{c}}\|_{1},
\end{align}
where $U=(K^{c}\cap E)\cup(K\cap E^{c})$. Since $d\geq1$ and $|U|=(1+\rho-2\alpha\rho)k\leq dk$, then $\|\bm{h}_{E}\|_{2}\leq\|\bm{h}_{\max(dk)}\|_{2}$ and $\|\bm{h}_{U}\|_{2}\leq\|\bm{h}_{\max(dk)}\|_{2}$, and thus clearly
\begin{align*}
w\|\bm{h}_{E}\|_{1}+(1-w)\|\bm{h}_{U}\|_{1}\leq w\sqrt{k}\|\bm{h}_{E}\|_{2}+(1-w)\sqrt{(1+\rho-2\alpha\rho)k}\|\bm{h}_{U}\|_{2}\leq\theta\sqrt{k}\|\bm{h}_{\max(dk)}\|_{2},
\end{align*}
where
\begin{align}\label{theta-defn}
\theta=w+(1-w)\sqrt{1+\rho-2\alpha\rho}.
\end{align}
This directly turns \eqref{new-added-1} to be the following inequality
\begin{align}\label{new-added-2}
\frac{\text{RHS}}{2\lambda}\leq2\left(w\|\bm{\widehat{x}}_{E^{c}}\|_{1}+(1-w)\|\bm{\widehat{x}}_{K^{c}\cap E^{c}}\|_{1}\right)+\theta\sqrt{k}\|\bm{h}_{\max(dk)}\|_{2}-\|\bm{h}_{E^{c}}\|_{1}.
\end{align}
Therefore, combing \eqref{Unconstrained-L1-L2Noise-Estimate-L} and \eqref{new-added-2} leads to the desired \eqref{NP-Lemma1-results1}, and \eqref{NP-Lemma1-results2} follows trivially from \eqref{NP-Lemma1-results1}.
\end{IEEEproof}

\begin{lemma}\label{NP-Lemma-RIP}
For any $g\geq1$£¬ if $A\in\mathbb{R}^{m\times n}$ satisfies the RIP of order $tk$ with RIC $\delta_{tk}$ and $t>g$, then for any vector $\bm{h}\in\mathbb{R}^{n}$ and any subset $S\subset[n]$ with $|S|=gk$, it holds that
\begin{align}\label{Lemma2-results}
\|\bm{h}_{S}\|_{2}\leq\beta_{1}\|A\bm{h}\|_{2}+\frac{\beta_{2}}{\sqrt{(t-g)k}}\|\bm{h}_{S^{c}}\|_{1},
\end{align}
where
\begin{align*}
\beta_{1}=\frac{2}{(1-\delta_{tk})\sqrt{1+\delta_{tk}}} \text{~and~}\beta_{2}=\frac{\delta_{tk}}{\sqrt{1-(\delta_{tk})^{2}}}.
\end{align*}
\end{lemma}
\begin{remark}
It is easy to know from Lemma \ref{NP-Lemma-RIP} that both $\beta_{1}$ and $\beta_{2}$ are two monotone increasing functions on the variable $\delta_{tk}$. Therefore if one restricts $\delta_{tk}$ to \eqref{sharp-RIP-L1}, it will be clear that
%\begin{align}
%\beta_{1}<&\frac{2\left(t-d+\theta^{2}\right)^{\frac{3}{4}}}{\theta\sqrt{\sqrt{t-d+\theta^{2}}-\sqrt{t-d}}}\triangleq\beta_{1}^{\sharp},\label{beta1-bound}\\
%=&\frac{2}{\theta^{2}}\left(t-d+\theta^{2}\right)^{\frac{3}{4}}\sqrt{\sqrt{t-d+\theta^{2}}+\sqrt{t-d}}\triangleq\beta_{1}^{\sharp}, \nonumber\\
%\beta_{2}<&\frac{\sqrt{t-d}}{\theta}\label{beta2-bound}.
%\end{align}
\begin{align}
\beta_{1}<&\frac{2}{\theta^{2}}\left(t-d+\theta^{2}\right)^{\frac{3}{4}}\sqrt{\sqrt{t-d+\theta^{2}}+\sqrt{t-d}}\triangleq\beta_{1}^{\sharp},\label{beta1-bound}\\
\beta_{1}>&2,~\beta_{2}<\frac{\sqrt{t-d}}{\theta}\triangleq\beta_{2}^{\sharp},\label{beta2-bound}
\end{align}
and
\begin{align}
\frac{1}{\sqrt{t-d}-\theta\beta_{2}}<\frac{\theta^{-1}}{\sqrt{(t-d)/(t-d+\theta^{2})}-\delta_{tk}}.\label{beta2-bound-new}
\end{align}
\end{remark}

\begin{IEEEproof}
The proof mainly follows from that of \cite[Lemma 2]{BPDN-2018-Submitted}. We here only give some key steps.

\textbf{Step 1}: For a given $t>g$, we start with denoting
\begin{align*}
S_{1}=&\left\{i\in S^{c}: |(\bm{h}_{S^{c}})_{i}|>\frac{\|\bm{h}_{S^{c}}\|_{1}}{(t-g)k}\right\},\\
S_{2}=&\left\{i\in S^{c}: |(\bm{h}_{S^{c}})_{i}|\leq\frac{\|\bm{h}_{S^{c}}\|_{1}}{(t-g)k}\right\}.
\end{align*}

\textbf{Step 2}: Using the similar skills in \cite{BPDN-2018-Submitted}, one can prove
\begin{align}\label{Lemma2-results-Middle}
\|\bm{h}_{S\cup S_{1}}\|_{2}\leq\beta_{1}\|A\bm{h}\|_{2}+\frac{\beta_{2}}{\sqrt{t-g}}\|\bm{h}_{S^{c}}\|_{1}.
\end{align}

\textbf{Step 3}: Proving \eqref{Lemma2-results} by \eqref{Lemma2-results-Middle} and $\|\bm{h}_{S}\|_{2}\leq\|\bm{h}_{S\cup S_{1}}\|_{2}$ .

These three steps are sufficient to prove Lemma \ref{NP-Lemma-RIP} when $tk$ is an integer. When $tk$ is not an integer, we define $t'=\lceil tk\rceil/k$, then $t'k$ is an integer and $\delta_{tk}=\delta_{t'k}$. Obviously, Lemma \ref{NP-Lemma-RIP} still holds in such case. In summary, Lemma \ref{NP-Lemma-RIP} will hold no matter whether or not $tk$ is an integer.
\end{IEEEproof}

\section{Main Results}
\label{Main-results}

With preparations above, we now give the main results.
\begin{theorem}\label{Theorem-L2Noise-RIP}
Assume that $\bm{b}$ is observed via \eqref{b=ax+z} with $\|\bm{z}\|_{2}\leq\epsilon$ and $E$ is denoted by $E=\text{supp}(\bm{\widehat{x}}_{\max(k)})$. Let $K\subseteq[n]$ be defined as in Lemma \ref{NP-Lemma1}. If the measurement matrix $A$ satisfies
\begin{align}\label{sharp-RIP-L1}
\delta_{tk}<\sqrt{\frac{t-d}{t-d+\theta^{2}}},
\end{align}
where $d$ and $\theta$ are denoted by \eqref{d-defn} and \eqref{theta-defn}, respectively, then
\begin{align}\label{Theorem-RIP-results1}
\|\bm{x}^{\sharp}-\bm{\widehat{x}}\|_{2}\leq C_{1}(\beta_{1},\beta_{2})\left(w\|\bm{\widehat{x}}_{E^{c}}\|_{1}+(1-w)\|\bm{\widehat{x}}_{K^{c}\cap E^{c}}\|_{1}\right)+C_{2}(\beta_{1},\beta_{2}),
\end{align}
where $\bm{x}^{\sharp}$ is the optimal solution of \eqref{UnConstrained-WL1} and
\begin{align*}
C_{1}(\beta_{1},\beta_{2})=&\frac{2\sqrt{k}\beta_{1}f_{1}(\beta_{2})\lambda+2f_{2}(\beta_{2})\epsilon}{r\sqrt{k}(r-\theta\beta_{2})(\theta\sqrt{k}\beta_{1}\lambda+\epsilon)},\\
C_{2}(\beta_{1},\beta_{2})=&\frac{\sqrt{k}\beta_{1}f_{3}(\beta_{2})\lambda+\left(\sqrt{d}(\beta_{2})^{2}+f_{2}(\beta_{2})\right)\epsilon}{r_{1}\sqrt{k}(r-\theta\beta_{2})(\theta\sqrt{k}\beta_{1}\lambda+\epsilon)^{-1}\lambda},
\end{align*}
with $r=\sqrt{t-d}$, $r_{1}=\sqrt{t-1}$ and $f_{i}(\cdot)$ for $i=1,2,3$ being denoted by \eqref{f1}, \eqref{f2} and \eqref{f3}, respectively.
\end{theorem}

\begin{remark}[\textbf{Recovery Condition}]
The established condition \eqref{sharp-RIP-L1} coincides with the one obtained recently by Chen, et al. in \cite{CS-prior-SP-CHen-2018}, which has been proved to be sharp for the exactly sparse signal recovery under noise-free measurements. However, their goal was to recover the signal with PKSI using the constrained model. On the other hand, our condition \eqref{sharp-RIP-L1} in fact is not a simple extension of the one in \cite{CS-prior-SP-CHen-2018}, but is obtained in a totally different way. We refer the interested readers to \cite{CS-prior-SP-CHen-2018} for more detailed discussion on \eqref{sharp-RIP-L1} and its potential corollaries.
\end{remark}

\begin{remark}[\textbf{Error Estimate}]
It seems that the obtained error estimate \eqref{Theorem-RIP-results1} shows a bit complicated since it integrates $\lambda$ and $\epsilon$ together. In what follows, we provide three special cases of \eqref{Theorem-RIP-results1} by selecting some simple but meaningful $\lambda$'s and/or $\epsilon$'s.

\textbf{Case 1)}: Suppose that $\lambda=\epsilon(\neq0)$, then by using \eqref{beta1-bound}, \eqref{beta2-bound} and \eqref{beta2-bound-new} we can deduce directly from \eqref{Theorem-RIP-results1} that
\begin{align*}
\|\bm{x}^{\sharp}-\bm{\widehat{x}}\|_{2}\leq\widetilde{C}_{1}&\left(w\|\bm{\widehat{x}}_{E^{c}}\|_{1}\right.+\left.(1-w)\|\bm{\widehat{x}}_{K^{c}\cap E^{c}}\|_{1}\right)+\widetilde{C}_{2}\lambda
\end{align*}
and
\begin{align*}
\widetilde{C}_{1}\triangleq&\frac{2\sqrt{k}\beta_{1}f_{1}(\beta_{2})+2f_{2}(\beta_{2})}{r\sqrt{k}(r-\theta\beta_{2})(\theta\sqrt{k}\beta_{1}+1)}
\leq\frac{2\beta_{1}^{\sharp}f_{1}(\beta_{2}^{\sharp})+2f_{2}(\beta_{2}^{\sharp})}{2r\theta\sqrt{k}(r-\theta\beta_{2})}\leq\frac{\widetilde{C}_{3}/\sqrt{k}}{\sqrt{(t-d)/(t-d+\theta^{2})}-\delta_{tk}},\\
\widetilde{C}_{2}\triangleq&\frac{\sqrt{k}\beta_{1}f_{3}(\beta_{2})+\sqrt{d}(\beta_{2})^{2}+f_{2}(\beta_{2})}{r_{1}\sqrt{k}(r-\theta\beta_{2})(\theta\sqrt{k}\beta_{1}+1)^{-1}}
\leq\frac{\beta_{1}^{\sharp}f_{3}(\beta_{2}^{\star})+\sqrt{d}(\beta_{2}^{^\sharp})^{2}+f_{2}(\beta_{2}^{^\sharp})}{r_{1}(r-\theta\beta_{2})}\sqrt{k}(\theta\beta_{1}^{\sharp}+1)\\
\leq&\frac{\sqrt{k}\widetilde{C}_{4}}{\sqrt{(t-d)/(t-d+\theta^{2})}-\delta_{tk}},
\end{align*}
where
\begin{align*}
\widetilde{C}_{3}=&\frac{2\beta_{1}^{\sharp}f_{1}(\beta_{2}^{\sharp})+2f_{2}(\beta_{2}^{\sharp})}{2r\theta^{2}},~\beta_{2}^{\star}=\mathop{\arg\min}_{u\in\{0, \beta_{2}^{\sharp}\}}f_{3}(u),\\
\widetilde{C}_{4}=&\frac{(\theta\beta_{1}^{\sharp}+1)\left(\beta_{1}^{\sharp}f_{3}(\beta_{2}^{\star})+\sqrt{d}(\beta_{2}^{^\sharp})^{2}+f_{2}(\beta_{2}^{^\sharp})\right)}{r_{1}\theta}.
\end{align*}
This directly yields
\begin{align*}
\|\bm{x}^{\sharp}-\bm{\widehat{x}}\|_{2}\leq&\frac{1}{\sqrt{(t-d)/(t-d+\theta^{2})}-\delta_{tk}}\bigg(\frac{\widetilde{C}_{3}}{\sqrt{k}}\left(w\|\bm{\widehat{x}}_{E^{c}}\|_{1}+(1-w)\|\bm{\widehat{x}}_{K^{c}\cap E^{c}}\|_{1}\right)+\sqrt{k}\widetilde{C}_{4}\lambda\bigg).
\end{align*}
This new error estimate also coincides with the ones in \cite{BPDN-2015-Shen,BPDN-2018-Submitted,BPDN-2019-JCAM,CoherenceWang-2018-submitted} in form, which are induced for the unconstrained models. However, their results do not take PKSI into consideration.

\textbf{Case 2)}: Suppose that $\lambda=\epsilon/\sqrt{k}(\neq0)$. Similar to the above analysis in \textbf{Case 1}, we can also obtain that
\begin{align*}
\|\bm{x}^{\sharp}-\bm{\widehat{x}}\|_{2}\leq&\frac{1}{\sqrt{(t-d)/(t-d+\theta^{2})}-\delta_{tk}}\bigg(\frac{\widetilde{C}_{5}}{\sqrt{k}}\left(w\|\bm{\widehat{x}}_{E^{c}}\|_{1}+(1-w)\|\bm{\widehat{x}}_{K^{c}\cap E^{c}}\|_{1}\right)+\widetilde{C}_{4}\epsilon\bigg),
\end{align*}
where
\begin{align*}
\widetilde{C}_{5}=&\frac{2\beta_{1}^{\sharp}f_{1}(\beta_{2}^{\sharp})+2f_{2}(\beta_{2}^{\sharp})}{r\theta(2\theta+1)}.
\end{align*}
This result coincides with the ones induced for the traditional constrained models in form, see, e.g., \cite{RIP-Recovery-conditions-tk-1,CS-prior-TIT-2012,ince2013nonconvex}, which to some extent indicates theoretically that the unconstrained model \eqref{UnConstrained-WL1} and the constrained model \eqref{Constrained-WL1} are equivalence in robust recovery of any (sparse) signals with PKSI.

\textbf{Case 3)}: Suppose that $\epsilon=0$, i.e., $\bm{z}=0$. In such case, it is also easy to deduce from \eqref{Theorem-RIP-results1} that
\begin{align*}
\|\bm{x}^{\sharp}-\bm{\widehat{x}}\|_{2}\leq&\frac{1}{\sqrt{(t-d)/(t-d+\theta^{2})}-\delta_{tk}}\bigg(\frac{\widetilde{C}_{6}}{\sqrt{k}}\left(w\|\bm{\widehat{x}}_{E^{c}}\|_{1}+(1-w)\|\bm{\widehat{x}}_{K^{c}\cap E^{c}}\|_{1}\right)+\sqrt{k}\widetilde{C}_{7}\lambda\bigg),
\end{align*}
where
\begin{align*}
\widetilde{C}_{6}=\frac{2f_{1}(\beta_{2}^{\sharp})}{r\theta^{2}}\text{~and~}\widetilde{C}_{7}=\frac{(\beta_{1}^{\sharp})^{2}f_{3}(\beta_{2}^{\star})}{r_{1}}.
\end{align*}
According to the above error estimate, it seems impossible to exactly recover any sparse signals through \eqref{UnConstrained-WL1} in the absence of noise. However, if one sets the parameter $\lambda$ to be a sufficient small positive number, the error between $\bm{x}^{\sharp}$ and $\bm{\widehat{x}}$ will tend to depend only on the available PKSI of the original signal itself. On the other hand, from the viewpoint of non-uniform recovery \cite{CS-textbook}, it has been shown that under certain conditions, one can successfully recover some (specific) sparse signals from the BPDN, see, e.g., \cite{BPDN-2004}. This may brings possibility for \eqref{UnConstrained-WL1} to realize the exact recovery of some sparse signals with PKSI when certain conditions are satisfied. More discussion on non-uniform recovery is beyond the scope of this paper, and we refer the interested readers to \cite{CS-textbook} and \cite{Non-uniform-VS-uniform} for details.
\end{remark}

\begin{remark}
Due to the limited space, we can not discuss more on the obtained results in this paper. We refer the interested readers to the supplementary material for more discussion on the weight choice and its resultant performance analysis.
\end{remark}

\begin{IEEEproof}
We first denote $\bm{h}=\bm{x}^{\sharp}-\bm{\widehat{x}}$, $F=\text{supp}(\bm{h}_{\max(dk)})$, $r=\sqrt{t-d}$, and $\eta= w\|\bm{\widehat{x}}_{E^{c}}\|_{1}+(1-w)\|\bm{\widehat{x}}_{K^{c}\cap E^{c}}\|_{1}$. Then we have
\begin{align}\label{key-inequality-used}
\|\bm{h}_{F^{c}}\|_{1}\leq\|\bm{h}_{E^{c}}\|_{1},
\end{align}
and also know from Lemma \ref{NP-Lemma-RIP} (with $S=F$) that
\begin{align}\label{Lemma2-results-new-1}
\|\bm{h}_{F}\|_{2}\leq\beta_{1}\|A\bm{h}\|_{2}+\frac{\beta_{2}}{r\sqrt{k}}\|\bm{h}_{F^{c}}\|_{1}.
\end{align}
Besides, combing \eqref{NP-Lemma1-results2}, \eqref{key-inequality-used} and \eqref{Lemma2-results-new-1} directly yields
\begin{align}\label{He-new-estimate}
&\|\bm{h}_{F}\|_{2}\leq\frac{\beta_{2}}{r\sqrt{k}}\left(2\eta+\theta\sqrt{k}\|\bm{h}_{F}\|_{2}+\frac{\epsilon}{\lambda}\|A\bm{h}\|_{2}\right)\nonumber\\
&~~~~~~~~~~~~+\beta_{1}\|A\bm{h}\|_{2}\nonumber\\
&=\frac{r\sqrt{k}\beta_{1}\lambda+\beta_{2}\epsilon}{r\sqrt{k}\lambda}\|A\bm{h}\|_{2}+\frac{2\beta_{2}}{r\sqrt{k}}\eta+\frac{\theta\beta_{2}}{r}\|\bm{h}_{F}\|_{2}\nonumber\\
&\leq\frac{r\sqrt{k}\beta_{1}\lambda+\beta_{2}\epsilon}{\sqrt{k}(r-\theta\beta_{2})\lambda}\|A\bm{h}\|_{2}+\frac{2\beta_{2}}{\sqrt{k}(r-\theta\beta_{2})}\eta,
\end{align}
where we used the condition \eqref{sharp-RIP-L1} and thus
\begin{align}\label{RIp-condition-ineqn}
\frac{\theta\beta_{2}}{r}-1=\frac{\theta\delta_{tk}}{\sqrt{(1-(\delta_{tk})^{2})(t-d)}}-1<0
\end{align}
for the last inequality. Similarly, we can also deduce from \eqref{NP-Lemma1-results2}, \eqref{key-inequality-used} and \eqref{Lemma2-results-new-1} that
\begin{align*}%\label{Hec-new-estimate}
\|\bm{h}_{F^{c}}\|_{1}\leq&2\eta+\theta\sqrt{k}\|\bm{h}_{F}\|_{2}+\frac{\epsilon}{\lambda}\|A\bm{h}\|_{2}\nonumber\\
\leq&2\eta+\theta\sqrt{k}\left(\beta_{1}\|A\bm{h}\|_{2}+\frac{\beta_{2}}{r\sqrt{k}}\|\bm{h}_{F^{c}}\|_{1}\right)\nonumber\\
&+\frac{\epsilon}{\lambda}\|A\bm{h}\|_{2}\nonumber\\
\leq&\frac{r(\theta\sqrt{k}\beta_{1}\lambda+\epsilon)}{(r-\theta\beta_{2})\lambda}\|A\bm{h}\|_{2}+\frac{2r}{r-\theta\beta_{2}}\eta.
\end{align*}

On the other hand, let $G$ denote the index set of the $k$ largest entries of $\bm{h}_{F^{c}}$ in magnitude. Then we can know from Lemma \ref{NP-Lemma-RIP} (with $S=G$) and \cite[inequality (2.3)]{BPDN-2015-Shen} that
\begin{align}
\|\bm{h}_{G}\|_{2}&\leq\beta_{1}\|A\bm{h}\|_{2}+\frac{\beta_{2}}{r_{1}\sqrt{k}}\|\bm{h}_{G^{c}}\|_{1},\label{lemma2-G}\\
\|\bm{h}_{F^{c}}\|_{2}&\leq\|\bm{h}_{G}\|_{2}+\frac{\|\bm{h}_{F^{c}}\|_{1}}{2\sqrt{k}},\label{References}
\end{align}
where $r_{1}=\sqrt{t-1}$. Then using \eqref{He-new-estimate} and \eqref{lemma2-G}, we have
\begin{align}\label{hG}
&\|\bm{h}_{G}\|_{2}\leq\beta_{1}\|A\bm{h}\|_{2}+\frac{\beta_{2}}{r_{1}\sqrt{k}}\left(\|\bm{h}_{F}\|_{1}+\|\bm{h}_{F^{c}}\|_{1}\right)\nonumber\\
&\leq\frac{\sqrt{d}\beta_{2}}{r_{1}}\left(\frac{r\sqrt{k}\beta_{1}\lambda+\beta_{2}\epsilon}{\sqrt{k}(r-\theta\beta_{2})\lambda}\|A\bm{h}\|_{2}+\frac{2\beta_{2}}{\sqrt{k}(r-\theta\beta_{2})}\eta\right)\nonumber\\
&~~~+\beta_{1}\|A\bm{h}\|_{2}+\frac{\beta_{2}}{r_{1}\sqrt{k}}\|\bm{h}_{F^{c}}\|_{1}\nonumber\\
&=\frac{\sqrt{d}\beta_{2}(r\sqrt{k}\beta_{1}\lambda+\beta_{2}\epsilon)+r_{1}\sqrt{k}\beta_{1}(r-\theta\beta_{2})\lambda}{r_{1}\sqrt{k}(r-\theta\beta_{2})\lambda}\|A\bm{h}\|_{2}\nonumber\\
&~~~+\frac{2\sqrt{d}(\beta_{2})^{2}}{r_{1}\sqrt{k}(r-\theta\beta_{2})}\eta+\frac{\beta_{2}}{r_{1}\sqrt{k}}\|\bm{h}_{F^{c}}\|_{1},
\end{align}
where we used $\|\bm{h}_{F}\|_{1}\leq\sqrt{dk}\|\bm{h}_{F}\|_{2}$ in the first inequality.

Now we estimate the upper bound of $\|A\bm{h}\|_{2}$. We first know from \eqref{NP-Lemma1-results1}, \eqref{key-inequality-used} and \eqref{Lemma2-results-new-1} that
\begin{align*}
&\|A\bm{h}\|_{2}^{2}-2\epsilon\|A\bm{h}\|_{2}\leq2\theta\sqrt{k}\lambda\left(\beta_{1}\|A\bm{h}\|_{2}+\frac{\beta_{2}}{r\sqrt{k}}\|\bm{h}_{F^{c}}\|_{1}\right)\\
&~~~~+4\lambda\eta-2\lambda\|\bm{h}_{E^{c}}\|_{1}\\
&=2\theta\sqrt{k}\beta_{1}\lambda\|A\bm{h}\|_{2}+4\lambda\eta+\frac{2(\theta\beta_{2}-r)\lambda}{r}\|\bm{h}_{E^{c}}\|_{1},
\end{align*}
which is equal to
\begin{align}\label{new-added-3}
\begin{split}
\|A\bm{h}\|_{2}^{2}-2(\theta\sqrt{k}\beta_{1}\lambda+&\epsilon)\|A\bm{h}\|_{2}-4\lambda\eta\\
&\leq\frac{2(\theta\beta_{2}-r)\lambda}{r}\|\bm{h}_{E^{c}}\|_{1}.
\end{split}
\end{align}
Using \eqref{RIp-condition-ineqn} again, we can further deduce from \eqref{new-added-3} that
\begin{align*}%\label{Theorem-L2Noise-Proof-1}
\|A\bm{h}\|_{2}^{2}-2(\theta\sqrt{k}\beta_{1}\lambda+\epsilon)\|A\bm{h}\|_{2}-4\lambda\eta\leq0.
\end{align*}
This directly leads to
\begin{align}\label{Ah}
&\|A\bm{h}\|_{2}\leq(\theta\sqrt{k}\beta_{1}\lambda+\epsilon)+\sqrt{(\theta\sqrt{k}\beta_{1}\lambda+\epsilon)^{2}+4\lambda\eta}\nonumber\\
&\leq(\theta\sqrt{k}\beta_{1}\lambda+\epsilon)+\sqrt{\left(\theta\sqrt{k}\beta_{1}\lambda+\epsilon+\frac{2\lambda}{\theta\sqrt{k}\beta_{1}\lambda+\epsilon}\eta\right)^{2}}\nonumber\\
&=\frac{2\lambda}{\theta\sqrt{k}\beta_{1}\lambda+\epsilon}\eta+2(\theta\sqrt{k}\beta_{1}\lambda+\epsilon).
\end{align}

Based on \eqref{Ah}, we can give two new upper bound estimates for $\|\bm{h}_{F}\|_{2}$ and $\|\bm{h}_{F^{c}}\|_{1}$, respectively, i.e.,
\begin{align}
\|\bm{h}_{F}\|_{2}\leq&\frac{2\sqrt{k}\beta_{1}(r+\theta\beta_{2})\lambda+4\beta_{2}\epsilon}{\sqrt{k}(r-\theta\beta_{2})(\theta\sqrt{k}\beta_{1}\lambda+\epsilon)}\eta\nonumber\\
&+\frac{2(r\sqrt{k}\beta_{1}\lambda+\beta_{2}\epsilon)(\theta\sqrt{k}\beta_{1}\lambda+\epsilon)}{\sqrt{k}(r-\theta\beta_{2})\lambda},      \label{Hf-new}\\
\|\bm{h}_{F^{c}}\|_{1}\leq&\frac{4r}{r-\theta\beta_{2}}\eta+\frac{2r(\theta\sqrt{k}\beta_{1}\lambda+\epsilon)^{2}}{(r-\theta\beta_{2})\lambda}.  \label{Hfc-new}
\end{align}
Now combing \eqref{References} and \eqref{hG}, together with \eqref{Ah}-\eqref{Hfc-new}, we have
\begin{align*}
\|\bm{h}\|_{2}\leq&\|\bm{h}_{F}\|_{2}+\|\bm{h}_{F^{c}}\|_{2}\\
\leq&\|\bm{h}_{F}\|_{2}+\|\bm{h}_{G}\|_{2}+\frac{\|\bm{h}_{F^{c}}\|_{1}}{2\sqrt{k}}\\
\leq&\frac{\sqrt{d}\beta_{2}(r\sqrt{k}\beta_{1}\lambda+\beta_{2}\epsilon)+r_{1}\sqrt{k}\beta_{1}(r-\theta\beta_{2})\lambda}{r_{1}\sqrt{k}(r-\theta\beta_{2})\lambda}\|A\bm{h}\|_{2}\\
&+\frac{2\sqrt{d}(\beta_{2})^{2}}{r_{1}\sqrt{k}(r-\theta\beta_{2})}\eta+\|\bm{h}_{F}\|_{2}+\frac{r_{1}+2\beta_{2}}{2r_{1}\sqrt{k}}\|\bm{h}_{F^{c}}\|_{1}\\
\leq&\frac{2\sqrt{k}\beta_{1}f_{1}(\beta_{2})\lambda+2f_{2}(\beta_{2})\epsilon}{r\sqrt{k}(r-\theta\beta_{2})(\theta\sqrt{k}\beta_{1}\lambda+\epsilon)}\eta\\
&+\frac{\sqrt{k}\beta_{1}f_{3}(\beta_{2})\lambda+\left(\sqrt{d}(\beta_{2})^{2}+f_{2}(\beta_{2})\right)\epsilon}{r_{1}\sqrt{k}(r-\theta\beta_{2})(\theta\sqrt{k}\beta_{1}\lambda+\epsilon)^{-1}\lambda},
\end{align*}
where
\begin{align}
f_{1}(\beta_{2})&=\theta\sqrt{d}(\beta_{2})^{2}+r(\sqrt{d}+2\theta)\beta_{2}+(2+\theta)rr_{1}, \label{f1}\\
f_{2}(\beta_{2})&=\sqrt{d}(\beta_{2})^{2}+2(r+r_{1})\beta_{2}+rr_{1},\label{f2}\\
f_{3}(\beta_{2})&=2\left(r\sqrt{d}-\theta(r_{1}-r)\right)\beta_{2}+(4+\theta)rr_{1}.\label{f3}
\end{align}
This completes the proof.
\end{IEEEproof}

\section{Conclusion}
\label{conclusion-part}
This paper aims to provide a deterministic (non-stochastic) analysis for the sparse recovery of signals with partially known support information from the weighted BPDN. Equipped with the powerful RIC notation, we established a series of sufficient conditions and their resultant error estimates. These theoretical results, to some degree, are well complementary for the recent ones of the weighted BPDN established in a stochastic manner.

%\appendices
%\section{Proof of Lemma}
%

\ifCLASSOPTIONcaptionsoff
  \newpage
\fi

\end{document}